# Continuous and reversible tuning of the disorder-driven superconductor–insulator transition in bilayer graphene

Gil-Ho Lee[†1], Dongchan Jeong[‡1], Kee-Su Park[2], Yigal Meir[3], Min-Chul Cha[§4], and Hu-Jong Lee[§1]

[1]Department of Physics, Pohang University of Science and Technology, Pohang 790-784, Republic of Korea

[2]Department of Physics, Sungkyunkwan University, Suwon 440-746, Republic of Korea

[3]Department of Physics, Ben-Gurion University of the Negev, Beer Sheva 84105, Israel

[4]Department of Applied Physics, Hanyang University, Ansan 426-791, Republic of Korea

[†]Present address: Department of Physics, Harvard University, Cambridge, MA 02138, USA

[‡]Present address: Semiconductor R&D Center, Samsung Electronics Co. LTD., Hwasung 445-701, Republic of Korea

[§]Correspondence and requests for materials should be addressed to M.-C.C. (email: mccha@hanyang.ac.kr) or to H.-J.L. (email: hjlee@postech.ac.kr).

# ABSTRACT

The influence of static disorder on a quantum phase transition (QPT) is a fundamental issue in condensed matter physics. As a prototypical example of a disorder-tuned QPT, the superconductor–insulator transition (SIT) has been investigated intensively over the past three decades, but as yet without a general consensus on its nature. A key element is good control of disorder. Here, we present an experimental study of the SIT based on precise *in-situ* tuning of disorder in dual-gated bilayer graphene proximity-coupled to two superconducting electrodes through electrical and reversible control of the band gap and the charge carrier density. In the presence of a static disorder potential, Andreev-paired carriers formed close to the Fermi level in bilayer graphene constitute a randomly distributed network of proximity-induced superconducting puddles. The landscape of the network was easily tuned by electrical gating to induce percolative clusters at the onset of superconductivity. This is evidenced by scaling behavior consistent with the classical percolation in transport measurements. At lower temperatures, the solely electrical tuning of the disorder-induced landscape enables us to observe, for the first time, a crossover from classical to quantum percolation in a single device, which elucidates how thermal dephasing engages in separating the two regimes.

The superconductor–insulator transition[1,2] (SIT) in disordered two-dimensional (2D) superconductors exhibits a zero-temperature separatrix between the superconducting and insulating phases[3,4]. The associated scaling behavior[5,6] reveals the intrinsic nature of the quantum criticality. Cooper pairs exist even in the insulating phase, which is evidenced by direct observations of the superconducting gap[7,8], as well as the earlier observation of a giant magnetoresistance[4,9], magneto-oscillations[10], and superfluid correlations[11]. These observations strongly suggest that the loss of pair coherence due to disorder drives the SIT. Several underlying mechanisms for the SIT have been suggested[1], including the dirty boson picture based on Anderson localization of Cooper pairs[12], classical[13] and quantum percolation[14] of superconducting clusters.

In the dirty boson model[5,12], which assumes that fluctuations of the pair amplitude are negligible, phase fluctuation of the superconducting order parameter destroys global superconductivity. However, some recent reports have pointed out that strong disorder induces amplitude fluctuations to form superconducting islands in the insulating phases, even in homogenously disordered thin films[7,15-18] (which is also relevant to the Higgs amplitude mode[19,20]). These amplitude fluctuations may drive the universality class of the SIT from the disordered boson class to a percolation universality class, governed by the loss of global connection of disordered superconducting islands.

This issue, concerning the interplay between disorder and superconductivity, is underscored by recent experiments, which have reported classical or quantum percolation critical behavior at the SIT in systems with varying degrees of disorder[21,22]. The relevance of these two different transitions, classical and quantum percolation, which are governed by different critical exponents, is determined by the question[14] whether percolating clusters are formed between superconducting islands via either superconducting paths carrying phase-disrupting currents (classical percolation) or via coherent quantum-tunneling links (quantum percolation). Thus, precise control of disorder is crucial to elucidate the interplay between disorder and thermal dephasing, which is responsible for the classical-to-quantum crossover behaviors, and in particular, to differentiate the disorder-induced geometrical effects on the SIT from generic density modulations.

Varying the thickness[1,3] of or annealing[23] superconducting thin films has been adopted in previous experiments to change the level of disorder. However, this can result in variations in the carrier density as well as the disorder landscape in a non-controllable way. Electrostatic gating has also been employed for 2D superconducting systems as a means of controlling the carrier density while preserving the spatial disorder on an atomic scale in superconducting films[24], heterostructures of complex oxides[25], high-$T_c$ superconductor[26], and graphene[27]. Here, we utilize electrostatic gating for accurate and reversible tuning of the disorder-induced landscape at energies close to the Fermi level by modulating both the carrier density and the band gap independently, rather than simple carrier density modulation with an uncontrolled fixed disorder.

Compared with deeply buried 2D electronic systems of semiconducting heterostructures or oxide interfaces, graphene is more chemically inert and easily accessible using contact probes. However, the carriers are not strongly localized in monolayer graphene (MLG), even at the charge-neutral point (CNP), where the nominal carrier density vanishes. This is accounted for by the presence of electron–hole puddles[28,29] produced by the disorder potential arising from charged defects on the substrate and/or chemical residues introduced during device fabrication. Since MLG has zero band gap, sufficiently doped bipolar conducting puddles may touch each other [Fig. 1(a)], making the boundaries transmissible by carriers via Klein tunneling. In contrast, in bilayer graphene (BLG), a band gap $E_g$ opens when an electric field is applied perpendicular to the graphene, separating the charged puddles [Fig. 1(b)], and the transport behavior becomes percolative. This feature of BLG allows a high degree of independent control of both the band gap and the carrier density in a wide range[30], as shown in Fig. 1(c), and provides flexibility in designing novel devices with controlled conductive behavior by fine-tuning the distance between puddles. Normal percolative transport has been reported in 2D electron gas systems in the low-carrier-density regime[31], and in MLG nanoribbons with a finite band gap close to the CNP[32]. Similar behavior was observed in this study for the gapped BLG with normal electrodes (see Materials and Methods). As the Andreev-paired carriers were induced by the proximity effect in our dual-gated BLG device, the system precisely simulates a percolative SIT via the puddles of the pairs, the geometry of which is determined by disorder tuning at the Fermi level.

## Results

**Gate-control of superconducting and insulating states.** Figure 1(d) shows a schematic diagram of the configuration of the dual-gated BLG device. A pair of Pb superconducting electrodes was closely attached to a mechanically exfoliated BLG layer, which was sandwiched between the top and bottom gates (see Methods). A scanning electron microscopy image of the device is shown in Fig. 1(e), together with the measurement configuration. The distance ($L$) between electrodes is 0.46 μm and the width ($W$) of the BLG is 7.0 μm. The contact resistances between BLG and Pb electrodes were negligibly small compared to the zero-bias junction resistance, $R$ (Supplemental Information, section 1 and Fig. S1). The BLG became superconducting as Andreev-paired carriers formed due to the proximity effect of the superconducting electrodes, along with the consequent Josephson coupling between them[33-35].
The voltages of the bottom gate, $V_b$, and the top gate, $V_t$, induced displacement fields $D_b = \varepsilon_b(V_b - V_{b,0}) / d_b$ and $D_t = -\varepsilon_t (V_t - V_{t,0}) / d_t$, along the $\hat{z}$ direction, where $\varepsilon$'s are the dielectric constants, $d$'s are the thicknesses of the dielectric layers, and $V_{b,0}$ ($V_{t,0}$) is the charge-neutral voltage offset of the bottom (top) gate due to the initial doping. The difference $D_{density} = D_b - D_t$ controls the carrier density (or the chemical potential), while the average, $D_{gap} = (D_b + D_t) / 2$, breaks the inversion symmetry of the BLG, opening up a band gap[36] (Supplemental Information, section 2 and Fig. S3).

Figure 2(a) shows the square resistance of the junction, $R_{sq} = R \times (W / L)$, as a function of $D_{density}$ and $D_{gap}$ measured at a base temperature of $T = 50$ mK. The superconducting and insulating states, marked by black and red symbols, respectively, were determined from the current–voltage ($I$–$V$) characteristics at each set of $D_{density}$ and $D_{gap}$. The two phases are separated coincidently by the quantum resistance of Cooper pairs, $R_Q = h / 4e^2$ (green contour line) as observed in other systems. On the weakly insulating side, the system exhibited nonlinear insulating $I$–$V$ characteristics, as shown in Fig. 2(b), the zero-bias conductance of which is consistent with 2D Mott variable range hopping conduction, $G(T) \sim \exp[-(T^*/T)^{1/3}]$, where $T^*$ is a characteristic temperature (see Methods). On the superconducting side, $R_{sq}$ eventually vanished, and a dissipationless supercurrent branch emerged, as shown in Fig. 2(c), which resulted from the proximity Josephson coupling (Supplemental Information, section 3 and Fig. S4).

**Finite-size scaling analysis on the temperature-dependent behavior.** The temperature dependence of $R_{sq}$ at different $D_{density}$ ranging from insulating to superconducting phases is shown in Fig. 3(a). It shows no signs of the re-entrance or kink of the resistance at temperatures down to 50 mK, which was commonly observed in granular films. Below the crossover temperature $T_0$ denoted by the broken line, $R_{sq}$ saturated, presumably due to Joule heating of charge carriers. The shift of $T_0$ to lower temperatures when the heating was reduced (i.e., when $R_{sq}$ was smaller) is consistent with the Joule-heating interpretation. In Fig. 3(b), the curves of $R_{sq}$ vs $D_{density}$ at different temperatures converge on a single point (i.e., $D_{density,c} \sim -0.3$ Vnm$^{-1}$) with a corresponding critical square resistance of $R_{sq,c} \sim 1.1 R_Q$, which is close to the universal value predicted by the dirty boson model for a low dissipative system.

The SIT behavior is interpreted as a quantum phase transition (QPT), as confirmed by $R_{sq}$ vs $D_{density}$ data converging to a single finite-size scaling curve[5,6] of the form $R_{sq} = R_{sq,c} f(xT^{-1/\nu z})$ close to the critical point [Figs. 3(c) and (d)]. Here, $f$ is a scaling function and $x \equiv |D_{density} - D_{density,c}|$ or $x \equiv |D_{gap} - D_{gap,c}|$ is a tuning parameter. The correlation length exponent $\nu$ and the dynamical critical exponent $z$ characterize the universality class of the QPT. The data for $400 < T < 600$ mK exhibit the best collapse, with a critical-exponent product of $\nu z = 1.44$, which is close to the value $\nu_{cl} = 4/3$ for classical percolation in 2D[37] [an exponent of $z = 1$ has been assumed for a system with charged particles[1], which also appears to be valid in our study, as found separately in the bias-field-tuned critical point]. However, at lower temperatures (i.e., $200 < T < 375$ mK), the best collapse was found with $\nu z = 2.59$, which is consistent with a quantum percolation transition in 2D with the value $\nu_q = 7/3$ (semi-classically one expects[38] $\nu_q = \nu_{cl} + 1$). The best estimates of $\nu_q$ in the literature[39] lie in the range 2.3-2.5. We will see below that these values were consistently found in several sweeps with different carrier densities and band gaps. Interestingly, there was a classical-to-quantum crossover at $T_1 \sim 400$ mK, which will be discussed later in a more quantitative manner. Theoretical studies have predicted[14,22] such a crossover from quantum to classical percolation due to decoherence at a finite temperature.

Observations of similar crossover behavior have been reported [22] for quantum Hall insulator transitions. However, no estimation was provided for the associated change in the electron temperature, $T_{el}$, introduced by the bias-induced Joule heating.

**Estimation of $T_{el}$ and the classical-to-quantum crossover.** Since Joule heating may seriously affect the behavior of the QPT, in particular, close to the lowest measurement temperature, we carried out an in-depth quantitative analysis of $T_{el}$. $T_{el}$ saturated to a temperature $T_0$ as the phonon temperature $T_{ph}$ (i.e., the measurement temperature) approached the base temperature, i.e., $T_{el} = T_0$ when $T_{ph} = 50$ mK. The dissipative power $P = I^2R$ at $T_{ph} = 50$ mK was estimated from the saturated resistance $R$ and the root-mean-square (r.m.s.) bias current of $I = 1$ nA, which exhibited a power-law dependence on $T_0$, as shown in Fig. 4(a), along with the fit to $P = A(T_{el}^\theta - T_{ph}^\theta) = A(T_0^\theta - T_{ph}^\theta)$ with $T_{ph} = 50$ mK. The fitting parameters were the electron–phonon coupling exponent $\theta = 2.8 \pm 0.1$ and the coefficient $A = 77 \pm 14$ fW·K$^{-2.8}$, where $T_0$ was estimated to be $T_0 = 160$ mK at the SIT point of $R_{sq} \sim 1.1R_Q$ (Supplemental Information, section 4 and Fig. S5). The value of $\theta$ was consistent with the recently observed value in MLG [40] in millikelvin regime.

With the electron temperature described by $T_{el} = [T_{ph}^\theta + T_0^\theta - (50 \text{ mK})^\theta]^{1/\theta} \approx (T_{ph}^\theta + T_0^\theta)^{1/\theta}$, we now discuss the temperature dependence of the critical exponents in detail. The exponent product $vz$ can be evaluated from the slope of a double logarithmic plot of $(dR/dx)_{x=0} \propto T^{-1/vz}$ vs $T$, as shown in Fig. 4(b), for each gate sweep of the $D_{density}$-tuned (sweep 1, 2, S1, and S2) and $D_{gap}$-tuned (sweep 3 and S3) SIT. Note that, in this plot, the heating effect is excluded by replacing the measurement temperature by the electron temperature with $T_0 = 160$ mK. For all gate sweeps, for $T > 400$ mK, the slope is described well by classical percolation

(red line), whereas for $T < 400$ mK, the slope is consistent with the quantum percolation model (blue line). Successful elimination of the Joule heating effect in this study made it possible to identify a crossover between classical and quantum percolation, with the temperature as a tuning parameter for the decoherence.

**Finite-size scaling analysis for bias electric field.** Similar to the temperature dependence, the bias current ($I$) dependence of $R_{sq}$ is also differentiated into two phases as shown in the inset of Fig. 5(a), such that $R_{sq}$ decreases (increases) with lowering $I$ in the superconducting (insulating) phase. Here, we emphasize that the value of $\theta$ satisfies the 'safety' criterion[6] $2/\theta > z/(z+1)$, where $z = 1$, for the intrinsic fluctuations being dominant in the Joule-heating effect. This allowed fitting of the critical-exponent product $\nu(z+1)$ for both the classical and quantum percolation regions from the scaling behavior as a function of the bias electric field. The finite-size scaling analysis with the electric field ($E$) in Figs. 5(a) and (b) provides additional information of $\nu(z+1)$, because the $E$ dependence of $R_{sq}$ has the form $R_{sq} = R_{sq,c}\ g[xE^{-1/\nu(z+1)}]$ near the critical point[6,41]. Here, $g$ is another scaling function. Similar to the $T$-varying scaling in Figs. 3(c) and (d), $E$-varying scaling also gives two different values of $\nu(z+1)$ depending on the bias current range. For $I = 9 - 15$ nA, the best scaling was obtained with $\nu(z+1) = 2.66$, which is close to the value of classical percolation [$\nu(z+1) = 8/3$]. But, for the lower bias current range of $I = 3 - 9$ nA the best fit was obtained with $\nu(z+1) = 4.56$, which is close to the value of quantum percolation [$\nu(z+1) = 14/3$].

**Products of critical exponents, $\nu z$ and $\nu(z+1)$.** $\nu(z+1)$, together with $\nu z$ from the $T$-varying scaling, allows an independent determination of the critical exponents[41] of $\nu$ and $z$. We investigated several critical points for both of $D_{density}$–driven and $D_{gap}$–driven SIT as indicated in Fig. 2(a). For each gate sweep, we performed scaling analysis for both the temperature and electrical field dependences to evaluate the critical-exponent products of $\nu z$ and $\nu(z+1)$, respectively. We summarized all the critical-exponent products in Figs. 5(c) and (d) for both of the classical and quantum percolation regimes. The corresponding scaling results for the classical percolation regime are shown in supporting information (Supplemental Information, section 5 and Figs. S7 and S8). At higher temperatures ($T > 400$ mK) or for higher electric fields ($I > 9$ nA), with the averaged values of $\nu z = 1.44 \pm 0.13$ and $\nu(z+1) = 2.81 \pm 0.31$ for all different gate sweeps (sweeps 1 – 3 and S1 – S3), we get $\nu = 1.37 \pm 0.34$ and $z = 1.05 \pm 0.27$. This result supports the SIT of charged bosons (Cooper pairs) in the classical percolation universality class, which is consistent with the percolative transport nature of carriers through charged puddles in BLG at $T = 4.2$ K. At lower temperatures ($T < 400$ mK) or lower electric fields ($I < 9$ nA), the averaged values of $\nu z = 2.83 \pm 0.33$ and $\nu(z+1) = 5.25 \pm 0.63$ give $\nu = 2.42 \pm 0.71$ and $z = 1.17 \pm 0.37$, which support the quantum percolation universality class for SIT.

**Discussion**

It is rather surprising that the BLG layer in the narrow region between the superconducting electrodes show the finite-size scaling behaviour of a 2D SIT, which is usually observed in homogeneous 2D systems. We believe that the temperature range of our transport measurements was sufficiently low as to allow the critical behaviour of the correlation length as a function of temperature. The observed temperature-dependent finite-size scaling was then governed by the temporal scale associated with the system temperature without apparent influence of the spatial scale of our device on the transition characteristics. Apparently, the spatial correlation length remained limited at finite temperatures (i.e., shorter than the spacing between the superconducting electrodes) as to neglect the effects arising from possible inhomogeneity of carrier transport or finite size of our system.

Our BLG devices provide a unique method to investigate the underlying mechanisms of SITs via accurate and reversible control of disorder. Electrical gating changed the average spacing between proximity-induced superconducting puddles to drive the QPT as Andreev-reflected bound pairs at the Fermi level establish long-range coherence via percolative paths to yield the critical power-law behavior of percolation with negligible thermal intervention. At lower temperatures than the range of classical percolation behavior, direct control of the disorder enabled us to estimate the effective electron temperature and consequently to identify the crossover between classical and quantum percolation in a single device. Previously, these two regimes have only been obtained in separate systems belonging to weak and strong disorder regimes[21]. Our proximity-coupled BLG system demonstrates that it is an exceptionally useful platform to study disorder-induced QPTs.

## Methods

**Device fabrication.** Fabrication of the bilayer-graphene Josephson-junction devices relied on mechanical exfoliation of graphene[42] on a highly doped silicon substrate, which was capped with a 300-nm-thick silicon oxide layer to form a bottom gate dielectric ($d_b$ = 300 nm, $\varepsilon_b$ = 3.9). Bilayer graphene was identified via optical contrast (Supplemental Information, section 6 and Fig. S9). Superconducting electrodes were defined using electron beam lithography and thermal evaporation of $Pb_{0.9}In_{0.1}$ onto the bilayer graphene. Indium was added to minimize the grain size and the surface roughness[35]. The junction area was covered with cross-linked poly(methyl methacrylate) (PMMA)[43,44], which formed a dielectric layer for the top gate ($d_t \approx 43$ nm, $\varepsilon_t =$ 4.5). A Ti/Au top-gate electrode stack (where the layers were 5- and 145-nm-thick, respectively) was deposited and accurately aligned to cover most (~90 %) of the junction area (Supplemental Information, Fig. S2). This allowed uniform gate control over the entire junction area. The thickness of the top gate dielectric, $d_t$, was determined from the shift of the resistance maximum of $V_t$ by the modulation of $V_b$. $V_{b,0}$ and $V_{t,0}$ were determined by comparing the band gap, which was estimated from the temperature dependence in Fig. 6.

**Low-noise measurements.** The sample was maintained in thermal contact with the mixing chamber of a dilution fridge (Oxford Kelvinox AST) and cooled to a base temperature of 50 mK. Electrical measurement lines were filtered by a combination of two-stage low-pass RC filters (with a cut-off frequency of ~ 30 kHz) mounted at the mixing chamber and pi-filters (with a cut-off frequency of ~ 10 MHz), which were at room temperature. We used a conventional lock-in technique with a bias current amplitude 1 nA r.m.s. at a frequency of 13.33 Hz for the temperature-dependent measurements, and a direct-current bias for the bias-field-dependent measurements.

**Temperature dependence of conductance at CNP.** At the charge neutrality point ($D_{density} = 0$), the Fermi level is placed in the middle of the bandgap $E_g$. Then, the conduction occurs with thermally activated carriers, providing the temperature ($T$) dependence of conductance, $G_{TA}(T) = G_{TA,0} \exp(-E_g/2k_BT)$, with Boltzmann constant $k_B$. However, in disordered bilayer graphene, bandgap is filled with the localized states such as conducting electron and hole puddles so that the carriers can hop across these states. Hopping transport is more pronounced at lower temperatures where the thermal activation (TA) is exponentially suppressed. As shown in Fig. 6(a), low-temperature conductance agrees with variable range hopping (VRH) model in two dimensions, $G_{VRH}(T) = G_{VRH,0} \exp[-(T_h/T)^{1/3}]$, whereas high-temperature data agree with the TA conduction. The measurement was done at temperatures above ~ 7 K, with the Pb electrodes in the normal state. The charge neutrality point for the top gate was estimated to be $V_{t,0} = -6.0$ V, where the $|D_{gap}|$ dependence of resultant fitting parameter $E_g$ agrees with the theoretical prediction of self-consistent tight-binding calculation as shown in Fig. 6(b). Similar TA+VRH transport properties were experimentally investigated in dual-gated bilayer graphene[45]. We could not directly determine $V_{t,0}$ as it was beyond the charge-leakage voltage of the top gate. However, the uncertainty in the determination of $V_{t,0}$ gives additional offsets to $D_{gap}$ only but does not affect the scaling analysis discussed in the text.

**Percolation transport in gapped bilayer graphene.** Carrier density inhomogeneity in two-dimensional (2D) GaAs semiconducting systems induces the percolative metal–insulator transition (MIT) in the low carrier density regime[31,46]. Similarly, graphene which has inhomogeneous charge puddles is also expected to exhibit the percolative MIT if a bandgap is introduced to separate the electron band from the hole band. For example, S. Adam *et al.*[32] fabricated graphene into a nanoribbon structure to open a bandgap in graphene and demonstrated a 2D MIT of the classical percolation universality class. There is also theoretical prediction of percolation behavior for bilayer graphene with a finite bandgap[47]. In our case, a vertical electric field opened a bandgap in bilayer graphene. We investigated transport properties of bilayer graphene in the presence of the superconducting proximity effect and analyzed them in the frame of percolative superconductor–insulator transition. To support the percolative transport characteristics in gapped bilayer graphene in the absence of superconductivity, we fabricated and performed control experiments with a device consisting of dual-gated bilayer graphene in contact with non-superconducting Ti/Au electrodes. Optical image of the device and the measurement configuration are shown in Fig. 7(a). While injecting current ($I$ = 1 nA r.m.s.) from $I^+$ to $I^-$, voltage drop between $V^+$ and $V^-$ was measured as a function of bottom ($V_b$) and top ($V_t$) gate voltages at the base temperature of $T$ = 4.2 K. According to the definition of $D_{density}$ and $D_{gap}$, a resistance map is plotted as a function of $D_{density}$ and $D_{gap}$ in Fig. 7(b). $D_{density}$ represents the carrier density ($n$ = 5.52 x $10^{12}$ cm$^{-2}$ x $D_{density}$·V$^{-1}$nm) accumulated by the electrical gates, while $D_{gap}$ determines opening of bandgap ($E_g$) in the bilayer graphene. $D_{density}$ dependence of conductance ($G$) at a fixed $D_{gap}$ = – 0.8 V/nm [along the red line in Fig. 7(b)] is plotted in Fig. 7(c) on log-log scale. The bandgap is estimated to be $E_g$ ~ 90 meV according to the self-consistent tight-binding model[30,48]. There appears three transport regimes depending on the $D_{density}$ in both electron and hole sides. In a highly doped state ($|D_{density}|$ > 0.5 V/nm), Fermi level exceeds the bandgap ($|E_F|$ > 100 meV) so that the system is expected to be in the *Boltzmann transport regime*[49] where $G \propto n$. In the range of 0.1 V/nm < $|D_{density}|$ < 0.5 V/nm, best fits to the *critical behavior* $G \propto (n - n_c)^\delta$ give exponents $\delta^h$ = 1.25±0.02 in the hole side and $\delta^e$ = 1.25±0.05 in the electron side, where $n_c$ is the critical carrier density. They are close to the theoretical prediction $\delta$ =4/3 for 2D classical percolation universality class. Near the charge neutrality point, $|D_{density}|$ < 0.1 V/nm, $G$ deviates from the percolation behavior and does not converge to zero but *becomes saturated*. This is because electron and hole puddles remain conducting even though the average carrier density vanishes at $D_{density}$ = 0. Figure 7(d) shows the same data and corresponding fitting lines of Fig. 7(c) on linear scale. The linear relation between $G$ and $n$ in the Boltzmann transport regime (blue lines) and the crossover between percolation and Boltzmann transport regimes (arrows) are more pronounced.

**Acknowledgments**

This work was supported by the National Research Foundation (NRF) through the SRC Center for Topological Matter (Grant No. 2011-0030046 for HJL), the GFR Center for Advanced Soft Electronics (Grant No. 2012M3A6A5055728 for HJL), the Basic Science Researcher Program (Grant No. 2010-0012134 for MCC), the Max Planck POSTECH/KOREA Research Initiative

Program (Grant No. 2011-0031558 for KSP), and the CRI Program at SKKU (Grant No. 2012R1A3A2048816 for KSP), funded by the Ministry of Science, ICT and Future Planning. Work at BGU was supported by a grant from the Israel Science Foundation.

## Author Contributions

D.J., G.-H.L., and H.-J.L. conceived the idea for the project. G.-H.L. and D.J. fabricated the devices and carried out the experiments. All authors analyzed the data. M.-C.C., Y. M., and K.-S.P. provided theoretical consultation on the scaling analysis. M.-C.C and H.-J.L. supervised the project. All authors contributed to the discussion and writing the manuscript.

## Competing Financial Interests Statement

The authors declare no competing financial interests.

# Figures and Figure Legends

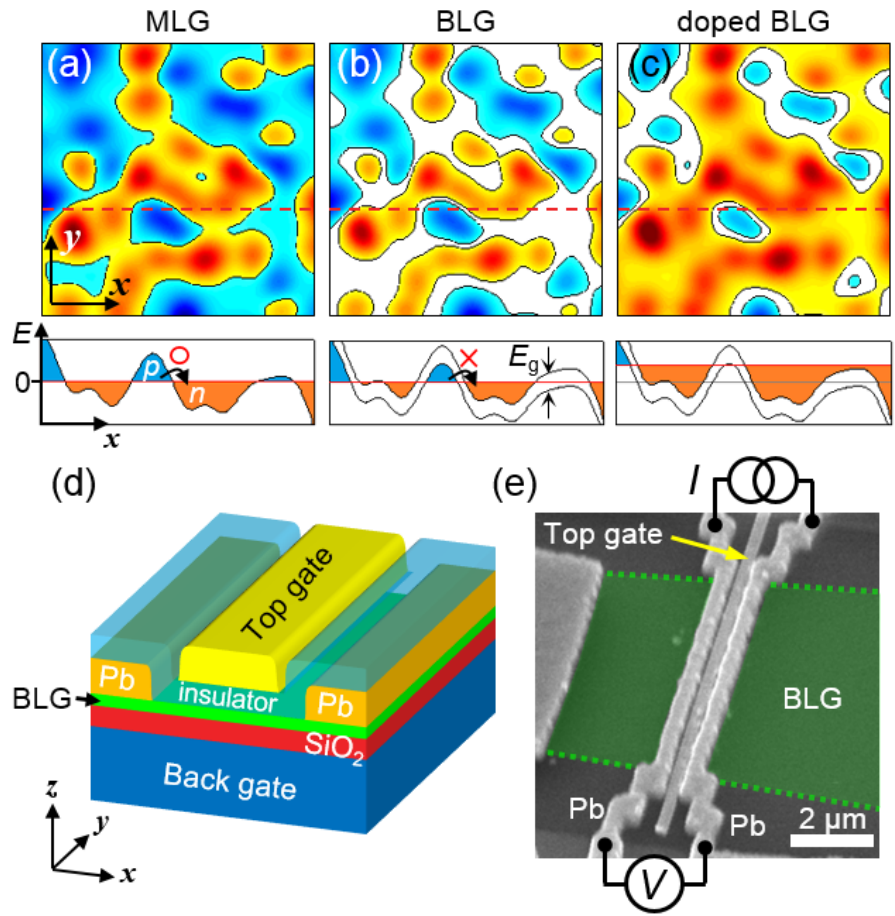

**Figure 1 | Dual-gated bilayer graphene Josephson junction.** (Upper panels) Spatially distributed charged puddles (a) in monolayer graphene (MLG) or zero-band-gap bilayer graphene (BLG), (b) BLG with a finite band gap $E_g$ in the charge-neutral state, and (c) $n$-doped finite-band-gap BLG. The red, blue, and white represent $n$- and $p$-doped (conducting) and finite-band-gap (insulating) states, respectively. (Lower panels) Cross-sections along the broken red curves in the upper panels, showing the variation in the conduction and valence bands. The solid red curve shows the chemical potential. (d) Schematic diagram showing the configuration of the dual-gated BLG Josephson junction. (e) Scanning electron microscopy image of the device, illustrating the measurement configuration. The green dotted lines show the location of the BLG.

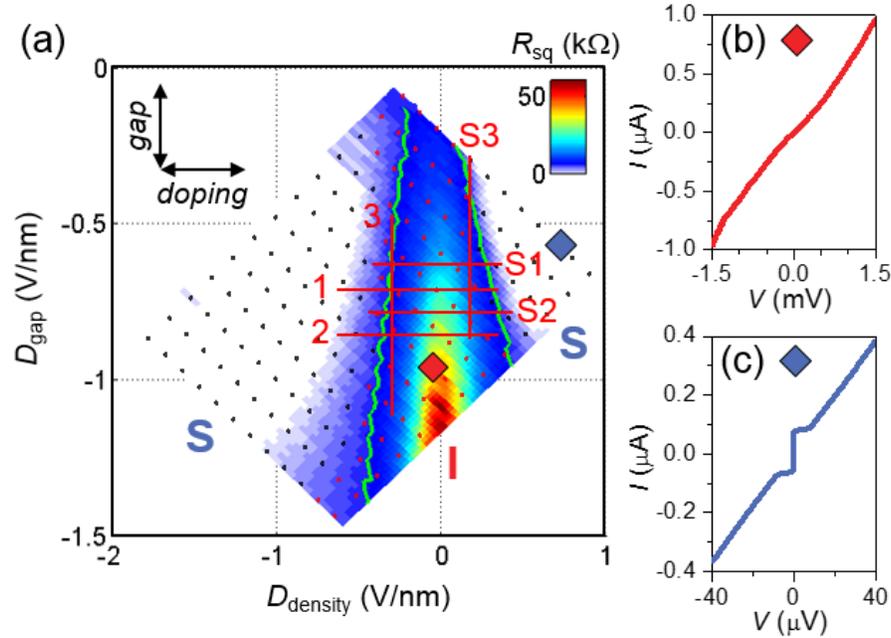

**Figure 2 | Insulating and superconducting states in an electron-hole puddle system.** (a) Color-coded square resistance map of the junction as a function of $D_{density}$ and $D_{gap}$ measured at 50 mK. The diamond symbols indicate representative insulating (red) or superconducting (blue) points of $D_{density}$ and $D_{gap}$. The green contours correspond to the quantum resistance of Cooper pairs, $R_Q = h/4e^2$, and separate insulating regions from the superconducting regions. The red curves indicate gate sweep traces 1–3 and S1–S3. Current–voltage characteristics taken at the corresponding points denoted in (a) exhibit (b) a nonlinear insulating behavior (the red curve) and (c) zero-resistance superconducting behavior (the blue curve).

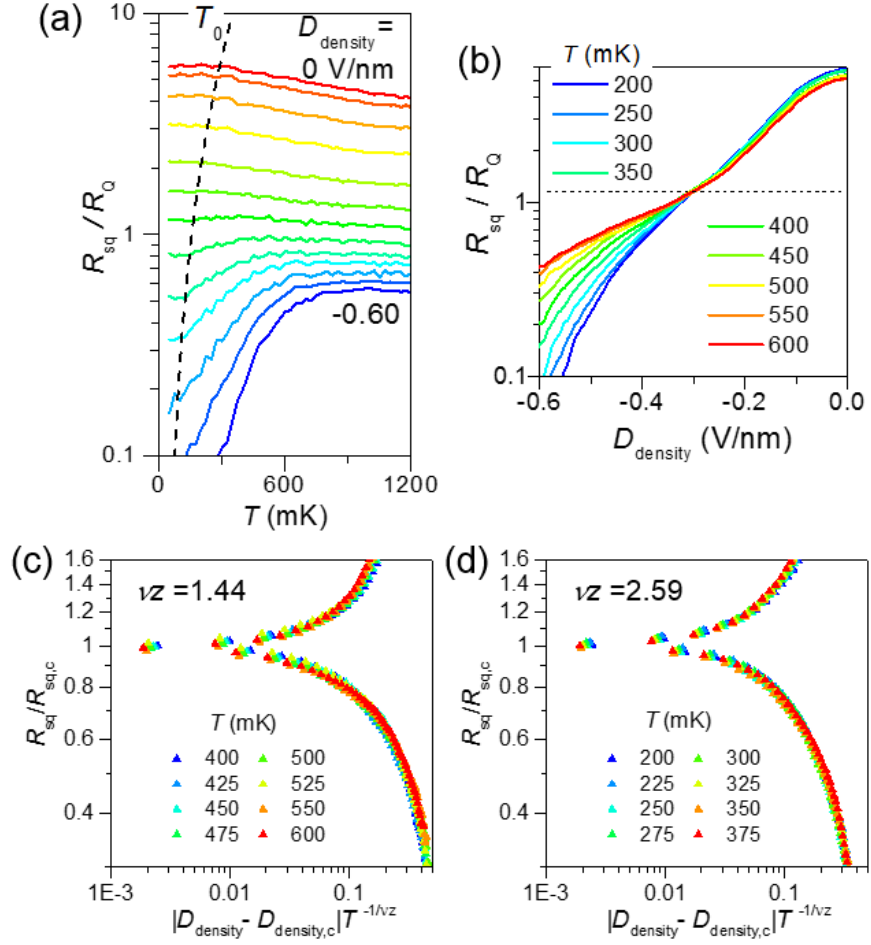

**Figure 3 | Temperature dependence of square resistance and its scaling behavior.** (a) Temperature dependence of the square resistance $R_{sq}$ with various $D_{density}$ for a fixed $D_{gap}$ = −0.86 Vnm$^{-1}$ plotted with a log-linear axis. $D_{density}$ was varied in steps of 0.05 Vnm$^{-1}$ from 0 Vnm$^{-1}$ (top) to −0.60 Vnm$^{-1}$ (bottom). The broken curve indicates the heating-induced crossover temperature $T_0$, below which the electron temperature and $R_{sq}$ saturated. (b) The data set in (a) plotted as a function of $D_{density}$ at various $T$. The horizontal broken line indicates the point of convergence. Finite-size scaling analysis of the $D_{density}$-driven superconductor–insulator transition for sweep 2 in Fig. 2(a). For 400 < $T$ < 600 mK (c), $\nu z$=1.44 gave the best data collapse; however, for temperatures of 200 < $T$ < 375 mK (d), $\nu z$ = 2.59 resulted in the best data collapse.

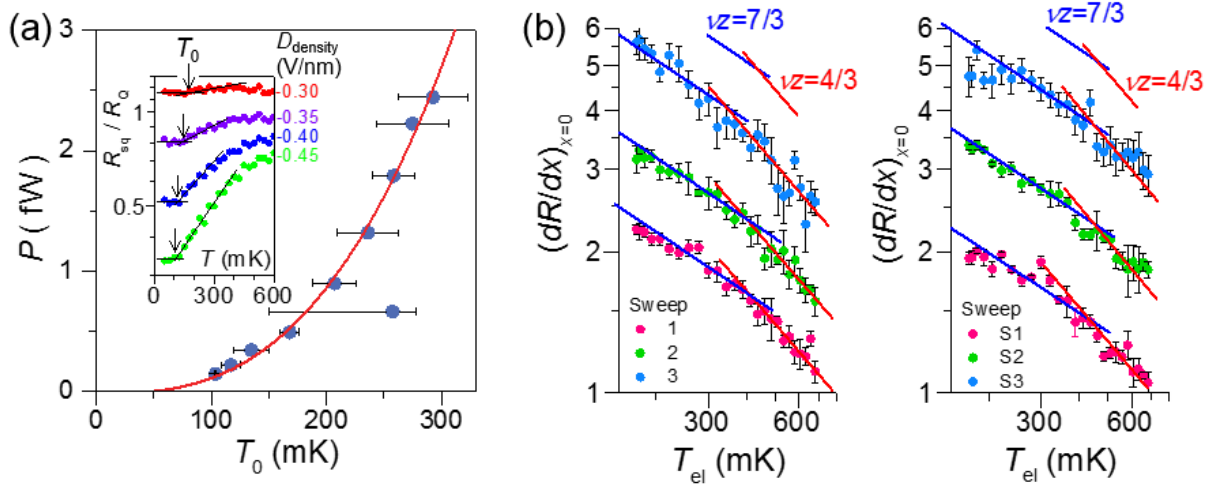

**Figure 4 | Estimation of electron temperature and temperature dependence of critical exponent products.** (a) Relationship between the heating-induced crossover temperature $T_0$ and the dissipative Joule-heating power $P$. The solid curve is a fit to $R_{sq}$ data close to the transition in a logarithmic scale. The horizontal lines are guides for the levelling-off of $R_{sq}$ and the arrows indicate $T_0$. (b) Electron temperature ($T_{el}$)-dependence of resistance slope $(dR/dx)_{x=0}$. For clarity, each data set was shifted vertically by an arbitrary offset. The red and blue linear curves indicate the slope expected with classical and quantum percolation, respectively.

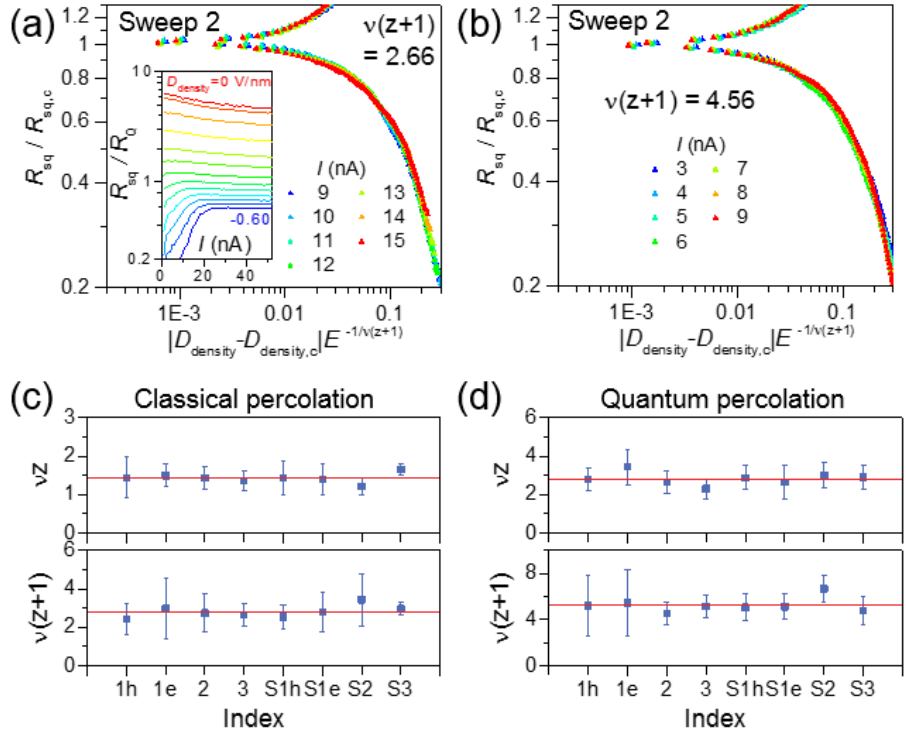

**Figure 5 | Scaling of square resistance for different electric fields and critical-exponent products for various gate sweeps.** Finite-size scaling analysis of electric field dependences of $D_{\text{density}}$-driven transition for the gate sweep 2, which yields the best collapse with $\nu(z+1) = 2.66$ for $I = 9 - 15$ nA in (a) or $\nu(z+1) = 4.56$ for $I = 3 - 9$ nA in (b). Inset of (a), bias current dependence of $R_{\text{sq}}$ at different doping levels ($D_{\text{density}}$, from 0 to $-0.60$ Vnm$^{-1}$, in steps of 0.05 Vnm$^{-1}$). Critical-exponent products $\nu z$ and $\nu(z+1)$ evaluated at various critical points for (c) the classical percolation regime and (d) the quantum percolation regime. Red lines represent average values. The characters 'h' and 'e' in the sweep indices stand for the hole and electron side, respectively.

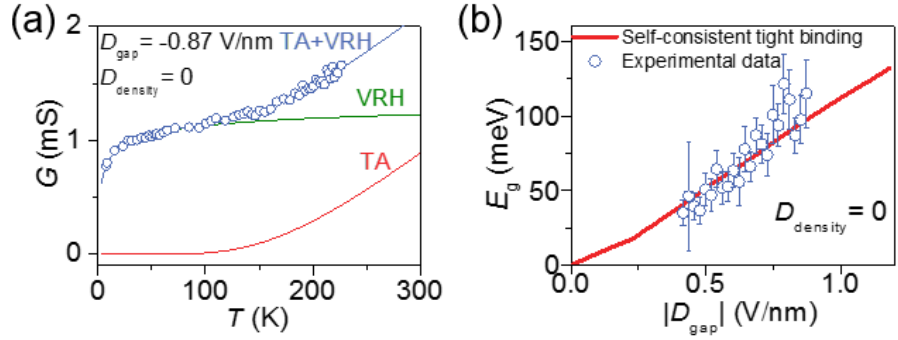

**Figure 6 | Thermal activation and variable range hopping conduction at the charge neutrality point.** (a) Temperature dependence of conductance $G$ at $D_{\text{gap}} = -0.87$ V/nm and $D_{\text{density}} = 0$ (charge neutrality point). The data were taken by ac measurements at zero dc bias and are fitted to the parallel conduction model (blue line), which consists of thermally activated conduction (red line) and variable-range-hopping conduction (green line). Best-fit parameters are $E_g = 115 \pm 23$ meV and $T_h = 1.92 \pm 0.59$ K. (b) Band gap $E_g$ estimated from the temperature dependence of conductance is plotted as a function of $|D_{\text{gap}}|$. Red line represents the prediction by self-consistent tight-binding calculation.

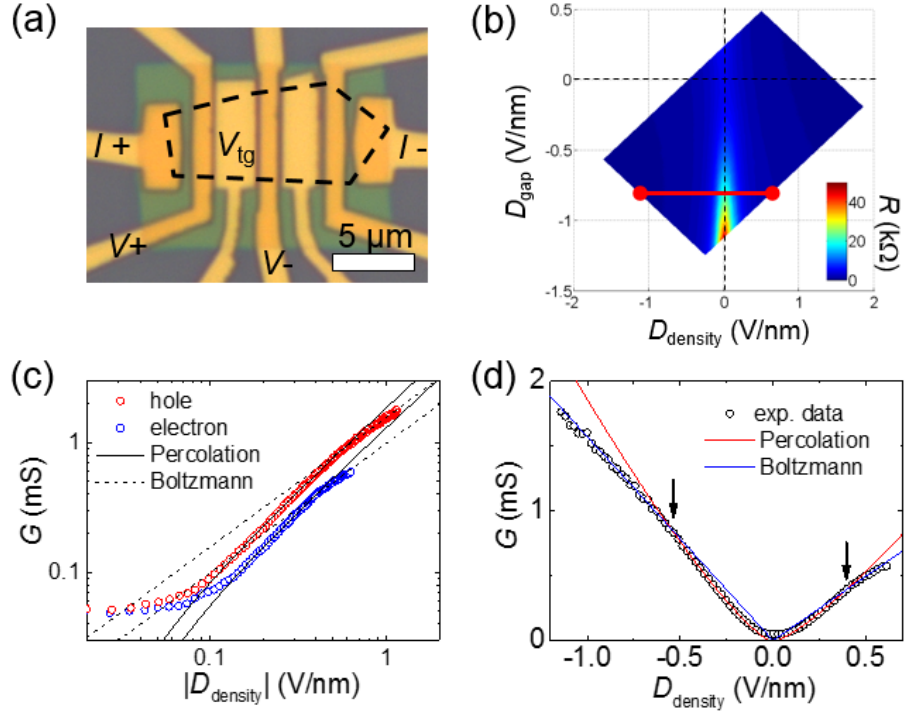

**Figure 7 | Percolation transport behavior in the gapped bilayer graphene.** (a) Optical image of the dual-gated bilayer graphene device with measurement configuration. (b) Resistance map as a function of $D_{density}$ and $D_{gap}$ measured at 4.2 K. (c) Log-log plot of conductance as a function of $|D_{density}|$ in the hole side (red symbols) and electron side (blue symbols). Best fits to the percolation behavior (solid lines) in the range of 0.1 V/nm < $|D_{density}|$ < 0.5 V/nm gives the critical exponent of $\delta^h = 1.25 \pm 0.02$ and the critical carrier density of $n_c^h = -1.00 \times 10^{11}$ cm$^{-2}$ in the hole side, and $\delta^e = 1.25 \pm 0.05$ and $n_c^e = 1.25 \times 10^{11}$ cm$^{-2}$ in the electron side. Dotted lines represent the Boltzmann transport behavior for highly doped state ($|D_{density}| > 0.5$ V/nm). (d) Linear plot of the same data and the fitting lines of (c). Arrows indicate the crossover between Boltzmann and percolation transport regimes.

# Supplementary Information

# Continuous and reversible tuning of the disorder-driven superconductor–insulator transition in bilayer graphene

Gil-Ho Lee[†1], Dongchan Jeong[‡1], Kee-Su Park[2], Yigal Meir[3], Min-Chul Cha[§4], and Hu-Jong Lee[§1]

[1]Department of Physics, Pohang University of Science and Technology, Pohang 790-784, Republic of Korea

[2]Department of Physics, Sungkyunkwan University, Suwon 440-746, Republic of Korea

[3]Department of Physics, Ben-Gurion University of the Negev, Beer Sheva 84105, Israel

[4]Department of Applied Physics, Hanyang University, Ansan 426-791, Republic of Korea

[†]Present address: Department of Physics, Harvard University, Cambridge, MA 02138, USA

[‡]Present address: Semiconductor R&D Center, Samsung Electronics Co. LTD., Hwasung 445-701, Republic of Korea

[§]Correspondence and requests for materials should be addressed to M.-C.C. (email: mccha@hanyang.ac.kr) or to H.-J.L. (email: hjlee@postech.ac.kr).

## 1. Contact resistance

Since the measured device resistance includes the contact resistance ($R_c$), we estimated $R_c$ using a four-probe measurement scheme shown in Fig. S1(a). The measured four-probe contact resistance [$R_{c,4p} = (V_+ - V_-)/I_{bias}$] of the left and right contacts were $-3$ $\Omega$ and $-4$ $\Omega$, respectively. Here, $V_+$ and $V_-$ are the electrical potential of the two electrodes and $I_{bias}$ is the bias current between $I_+$ and $I_-$ contact leads. The negative value of $R_{c,4p}$ in the cross-junction geometry can be understood as that $R_c$ is much smaller than the electrode resistance $R_{line}$ (~ 7 $\Omega$), resulting in non-uniform current flow along the junction #1. When $R_c$ is sufficiently larger than $R_{line}$, bias current flows uniformly through the junction along the vertical direction and each top and bottom electrode becomes equipotential. This results in voltage difference $\Delta V = V_+ - V_-$ to be positive [Fig. S1(b)] and $R_{c,4p}$ well represents $R_c$. However, when $R_c$ is sufficiently smaller than $R_{line}$, the electrode on top and the graphene layer at bottom behave as a single piece with the current flow becoming nonuniform along the junction. In this case, $V_-$ gets higher than $V_+$, which leads to a negative value of $R_{c,4p}$. This feature is confirmed in the numerical simulation for different values of $R_c$ in Fig. S1(c) and S1(d). Simulation was done by commercial package COMSOL Multiphysics with the same geometrical and electrical parameters of the device. As $R_c$ gets smaller than $R_{line}$, $R_{c,4p}$ becomes negative and saturated to the value of $-R_{line}$ [Fig. S1(e)], which is close to the experimentally measured value of $R_{c,4p}$. This ensures that $R_c$ of our device was an order of a few ohms, which was negligible compared to the device resistance (a few hundreds ohms).

## 2. Coordinate transformation of the resistance map

Experimentally, we constructed a resistance map as a function of the bottom gate and top gate voltages ($V_b$ and $V_t$, respectively) as shown in Fig. S3(a). The carrier doping and the gap opening of the bilayer graphene were exclusively determined by the parameters $D_{density}$ ($= D_b - D_t$) and $D_{gap}$ [$= (D_b + D_t) / 2$], respectively, with $D_b = \varepsilon_b(V_b - V_{b,0})/d_b$ and $D_t = -\varepsilon_t (V_t - V_{t,0})/d_t$. Here, $\varepsilon$ is the dielectric constant, $d$ is the thickness of dielectric layers, and $V_{b,0}$ ($V_{t,0}$) is the charge-neutral gate voltage of the bottom (top) gate due to the initial environmental doping. Thus, for convenience, we transformed the coordinate for the resistance map from the ($V_b$, $V_t$) basis system to the ($D_{density}$, $D_{gap}$) basis system as shown in Fig. S3(b).

## 3. Josephson coupling in the superconducting phase

When the square resistance becomes smaller than the quantum resistance, the superconducting phase emerges in the region of bilayer graphene layer. As discussed in the main text, the superconducting phase is induced by the proximity effect from the superconducting electrodes. In this section, we present the genuine Josephson coupling via the bilayer graphene layer, confirmed by microwave irradiation and applying perpendicular magnetic fields on the bilayer-graphene Josephson junction. When a microwave was irradiated on the Josephson junction, the beating of ac voltage and the ac Josephson effect generated equidistant voltage steps in the current−voltage characteristics [Fig. S4(a)], which is known as Shapiro steps[1]. In Fig. S4(b), the voltage step size $\Delta V$ shows highly linear relationship with the irradiated microwave frequency $f_{mw}$ as $\Delta V = h f_{mw}/2e$ with Planck's constant $h$ and electron charge $e$.

Microwave amplitude ($P^{1/2}$) dependence of differential resistance ($dV/dI$) with fixed $f_{mw}$ = 5 GHz is plotted in Fig. S4(c). Shapiro steps ($dV/dI = 0$) shows well-behaving Bessel-function-like oscillation as a function of $P^{1/2}$.

Another unique feature of Josephson junction is periodic oscillation of critical current ($I_c$) with applied perpendicular magnetic field ($B$), which is known as Fraunhofer pattern[1]. When the magnetic flux $\Phi = BA_{eff}$ threading the effective junction area $A_{eff}$ becomes an integer multiple of magnetic flux quantum $\Phi_0 = h/2e$, $I_c$ drops to zero except for B=0. Here, $A_{eff}=W(L+2\lambda_L)$ with taking into account of the London penetration depth $\lambda_L$ of the superconducting Pb-In electrodes. $W = 7.0$ μm is the width and $L = 0.46$ μm is the length of the Josephson junction. In Fig. S4(d), the $B$ dependence of $I_c$ clearly manifests Fraunhofer pattern with periods of $\Delta B \sim 2.8$ G, which agrees with the theoretical prediction of $\Delta B$ with $\lambda_L \sim 0.3$ μm obtained in the independent measurements[2].

## 4. Heat dissipation by electron-phonon coupling in bilayer graphene in low temperature regime

The saturation behaviour of resistance by the dissipative Joule heating shown in Figs. S5(a), (b), and (c) gives information about the electron-phonon coupling in the bilayer graphene Josephson junction device. Crossover temperature ($T_0$) and the saturation resistance ($R$) correspond to the electron temperature in association with the base sample holder temperature and the dissipative power $P = I^2R$, respectively, with bias current $I = 1$ nA r.m.s. Most of the heat generated by the bias current is dissipated via electron-phonon coupling, since hot electron diffusion into the electrodes can be ignored due to the exponentially suppressed quasiparticle density of states of the lead (Pb) superconducting electrode[3]. Also, we can assume that the phonon of bilayer graphene is fully thermalized to the temperature of the silicon oxide substrate since the interfacial thermal resistance is a few orders of magnitude smaller than the thermal resistance between electron and phonon of the bilayer graphene[3,4]. Here, the interfacial thermal resistance at low temperature is estimated by extrapolating the experimental data in Ref. [4].

Fig. S5(d) displays the relation between crossover temperature and $P$ along with the best-fit curve of $P = A(T_{el}^{\theta} - T_{ph}^{\theta})$, giving the best-fit value of electron-phonon coupling exponent $\theta = 2.8 \pm 0.1$ for the (base) phonon temperature $T_{ph} = 50$ mK and the coefficient $A = 77 \pm 14$ fW·K$^{-2.8}$. Here, we assumed that the electron temperature ($T_{el}$) at the base temperature is saturated to $T_0$. The exponent $\theta = 2.7 \pm 0.1$ was also determined by the slope in double logarithmic plot in Fig. S5(e), assuming that $T_{ph}^{\theta}$ term was negligible compared to $T_{el}^{\theta}$ for $T_{el} > 100$ mK. The exponent $\theta$ was smaller than 4 and close to 3, which mimicked the electron-phonon coupling in disordered monolayer graphene systems in millikelvin temperature range[3,5]. This low value of the exponent ($\theta < 4$) makes bilayer graphene system a unique platform for the bias-dependent finite-size scaling studies for the independent determination of a dynamical critical exponent. This sharply contrasts with ordinary two-dimensional electron systems[6] (with $\theta = 4 - 7$), which are easily driven into 'dangerous' regime where Joule heating significantly enhances the electron temperature and thus obscures the quantum critical scaling behaviour.

## 5. Finite-size scaling with bias electric field

As discussed in the main text, the bias current ($I$) dependence of $R_{sq}$ is also differentiated into two phases and enables finite-size analysis on electric field ($E$). Analysis similar to the one in Fig. 4(b) is adopted to determine the exponent $v(z+1)$, but now $(dR/dx)_{x=0}$ is plotted as a function of $I$ as shown in Fig. S6. Using $I$ instead of electric field $E$ ($\propto IR$) as an external parameter is valid because the resistance at SIT point is universal irrespective of $I$. The crossover from classical to quantum percolation with lowering $I$ resembles the previous observation in the $T$-dependent scaling. More scattering of the data for smaller $I$ is due to reduced signal-to-noise ratio in dc measurements. Here, one should be careful lest the Joule heating power $P$ ($\propto E^2$) should enhance the carrier temperature as $T_{el} \propto P^{1/\theta}$ and alter the intrinsic scaling behaviour. For the observed exponent $\theta = 2.8$ in BLG in the Section 5, our quantum critical scaling would be in the 'safety' criterion[7] of $2/\theta > z/(z+1)$ with $z=1$, where the self-heating effect was negligible compared with the intrinsic fluctuation effects.

## 6. Estimation of the number of graphene layers

To identify the number of graphene flakes, we used the intensity contrast in the green light range[8]. Fig. S6(a) shows the optical image of graphene flakes exfoliated on a highly electron-doped Si substrate capped with a 300-nm thick $SiO_2$ layer. Green light contrast ($C_{green}$) of the graphene flakes shows the linear relationship to the number of graphene layers as shown in Fig. S6(b). Bilayer graphene part (region 2) was selected to fabricate the dual-gated bilayer graphene Josephson junction device in this study.

**Supplementary References**

# Figures and Figure Legends

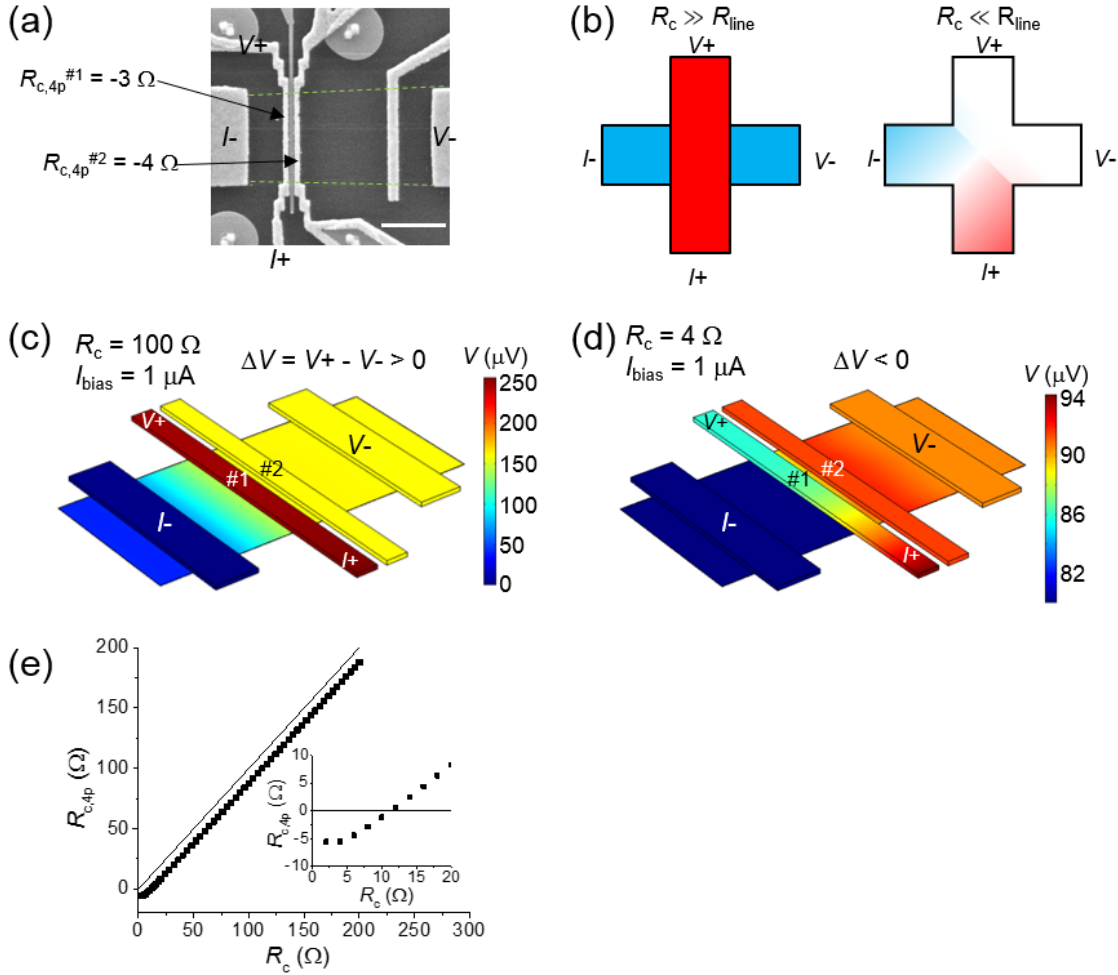

**Fig. S1.** Contact resistance. (a) Scanning electron microscope (SEM) image of the device with the contact-resistance measurement scheme. Green dotted lines denote the boundary of the bilayer graphene. (b) Schematics of voltage profile of high and low contact-resistance regimes. Red and blue colours represent high and low electric potential, respectively. (c, d) Numerical simulation of voltage profiles with contact resistance, $R_c = 100\ \Omega$ (c) and $R_c = 4\ \Omega$ (d). (e) Simulated four-probe contact resistance ($R_{c,4p}$) as a function of $R_c$. Solid line represents $R_{c,4p} = R_c$. Inset, a close-up view of (e). As $R_c$ gets lower than the electrode resistance $R_{line} \sim 7\ \Omega$, $R_{c,4p}$ can be negative.

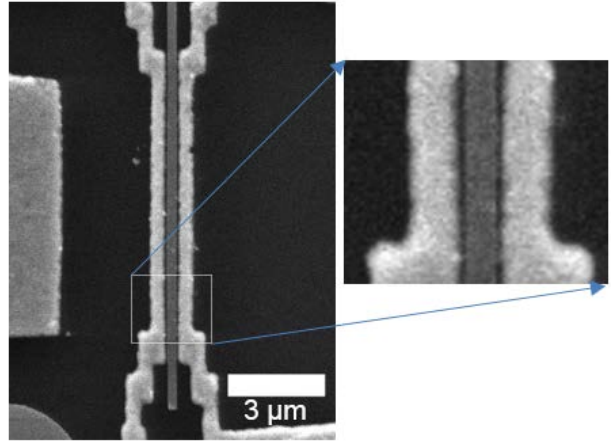

**Fig. S2.** Scanning electron microscope (SEM) image. SEM image of a dual-gated bilayer graphene Josephson junction device. Magnified image shows the top gate that is aligned to the superconducting electrodes as close as possible without touching them to avoid the gate leakage. The gap between the top gate and superconducting electrodes is less than 20 nm.

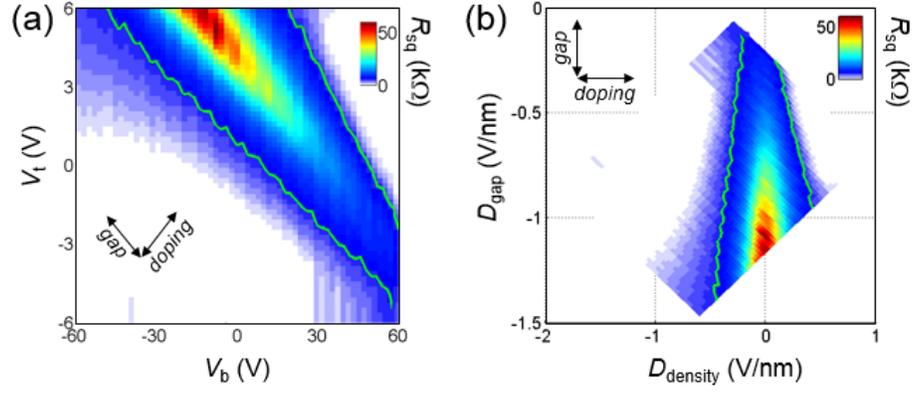

**Fig. S3.** Coordinate transformation. (a) Colour-coded plot of the junction resistance in ($V_b$, $V_t$) coordinate system. (b) The same in ($D_{density}$, $D_{gap}$) coordinate system. The green contour lines correspond to the quantum resistance of Cooper pairs, $R_Q = h/4e^2$.

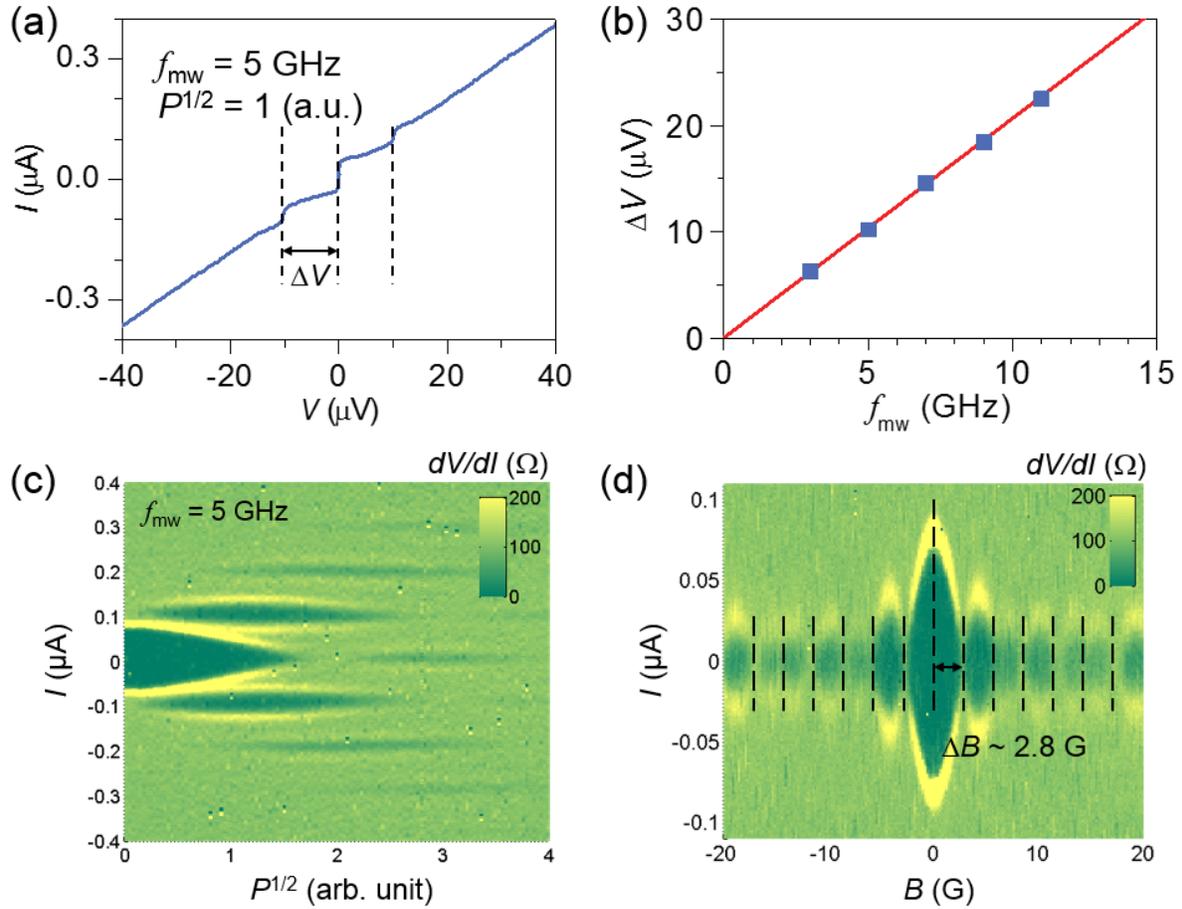

**Fig. S4.** Josephson coupling in the superconducting phase. (a) Equidistant voltage steps ($\Delta V$) appear in current−voltage characteristics with microwave irradiation of frequency $f_{mw}$ = 5 GHz. (b) $f_{mw}$ dependence of $\Delta V$ (symbols) agrees with theoretically predicted linear relationship, $\Delta V = hf_{mw}/2e$ (red line). (c) Microwave amplitude ($P^{1/2}$) dependence of Shapiro steps at a fixed frequency $f_{mw}$ = 5 GHz shows quasi-periodic Bessel-function-like oscillations. (d) Perpendicular magnetic field ($B$) dependence of the junction critical current shows the Fraunhofer pattern in constant periods of $\Delta B \sim 2.8$ G.

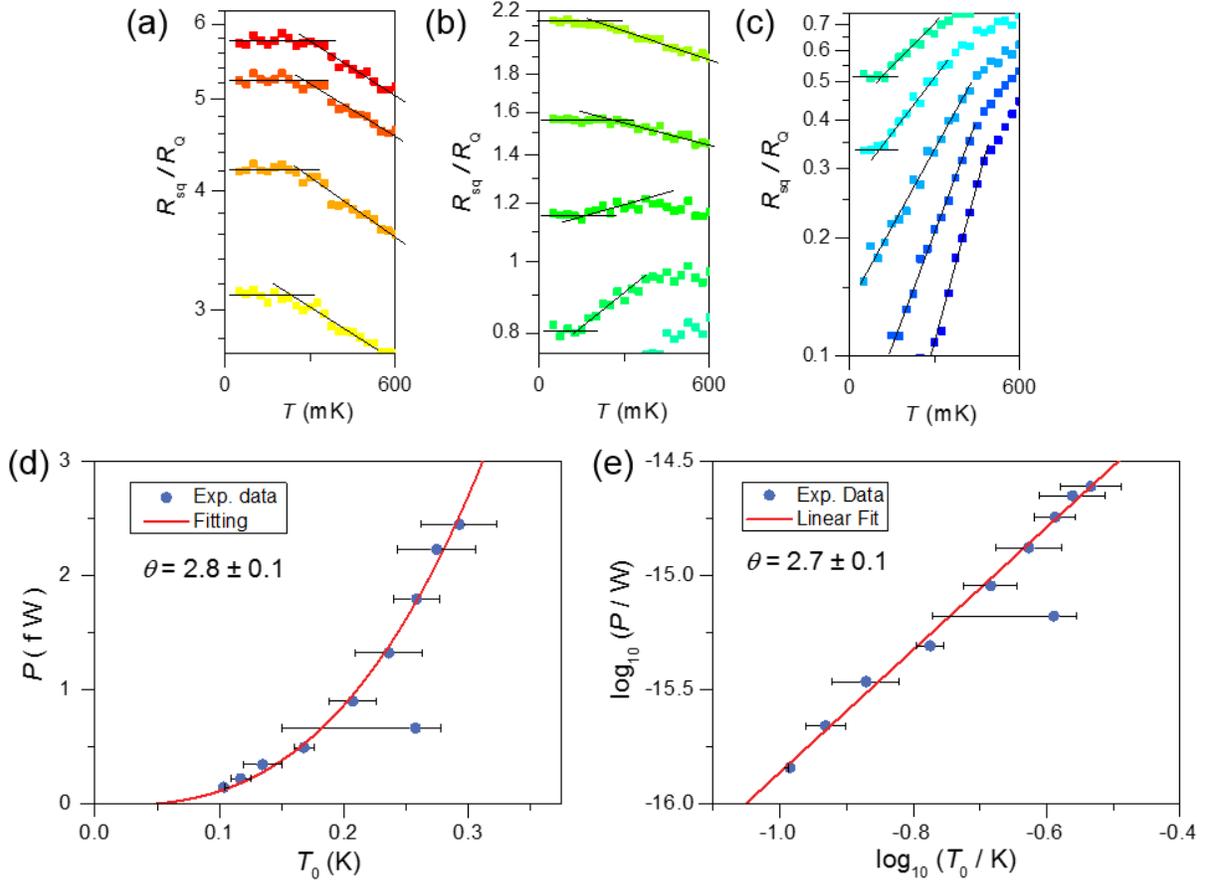

**Fig. S5.** Enhancement of electron temperature due to the dissipative Joule heating. Temperature dependence of square resistance ($R_{sq}$) divided by quantum resistance of Cooper pair ($R_Q$) plotted in a semi-log scale for a fixed $D_{gap} = -0.86$ Vnm$^{-1}$ (a) at $D_{density} = 0$ (top), $-0.05$, $-0.10$, $-0.15$ (bottom) Vnm$^{-1}$, (b) $D_{density} = -0.20$ (top), $-0.25$, $-0.30$, $-0.35$ (bottom) Vnm$^{-1}$, (c) $D_{density} = -0.40$ (top), $-0.45$, $-0.50$, $-0.55$, $-0.60$ (bottom) Vnm$^{-1}$. Saturation of resistance is guided by solid lines. (d) Relation between crossover temperature ($T_0$) and dissipative power by Joule heating ($P$). Solid line is the best-fit curve. (e) Log-log plot of $T_0$ versus $P$ and the corresponding the best linear fit.

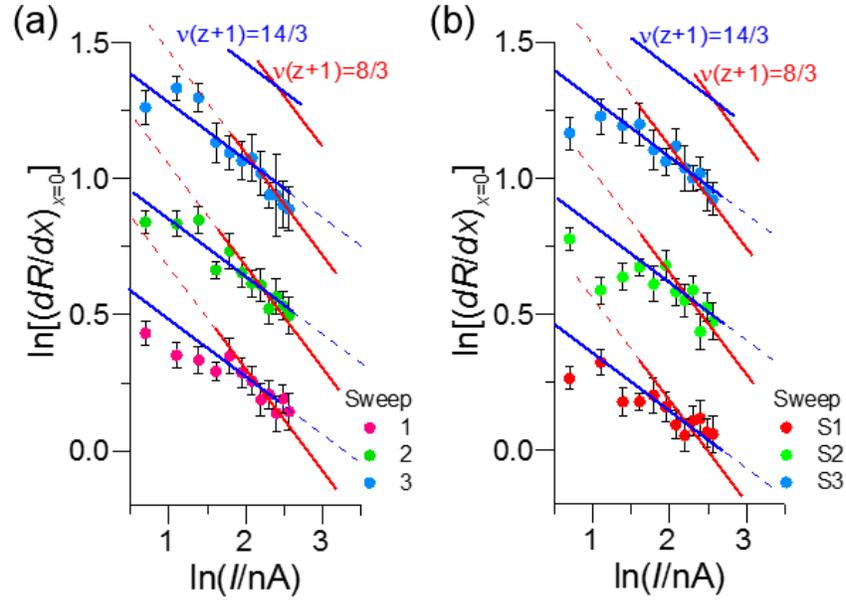

**Fig. S6.** Bias-current dependence of the resistance slope at the transition $(dR/dx)_{x=0}$ for different gate sweeps (a) 1 – 3 and (b) S1 – S3. Each set of data is plotted with an arbitrary vertical shift for clarity. Red (blue) straight line shows the expectation of the classical (quantum) percolation.

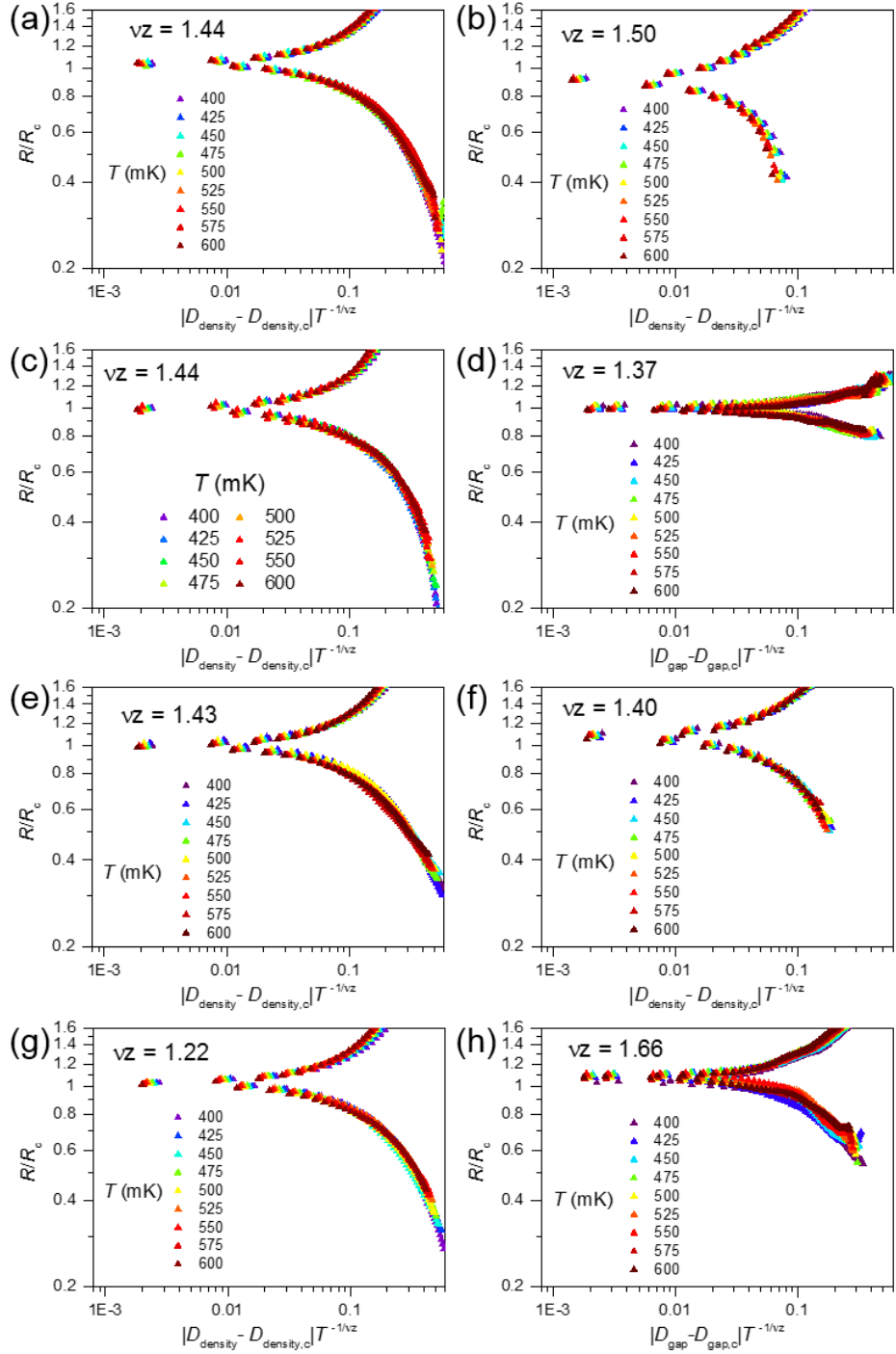

**Fig. S7.** Scaling analysis for temperature variation. For the sweep 1 (a) in the hole and (b) electron sides. For (c) the sweep 2 and (d) 3. For the sweep S1 (e) in the hole and (f) electron sides. For (g) the sweep S2 and (h) S3.

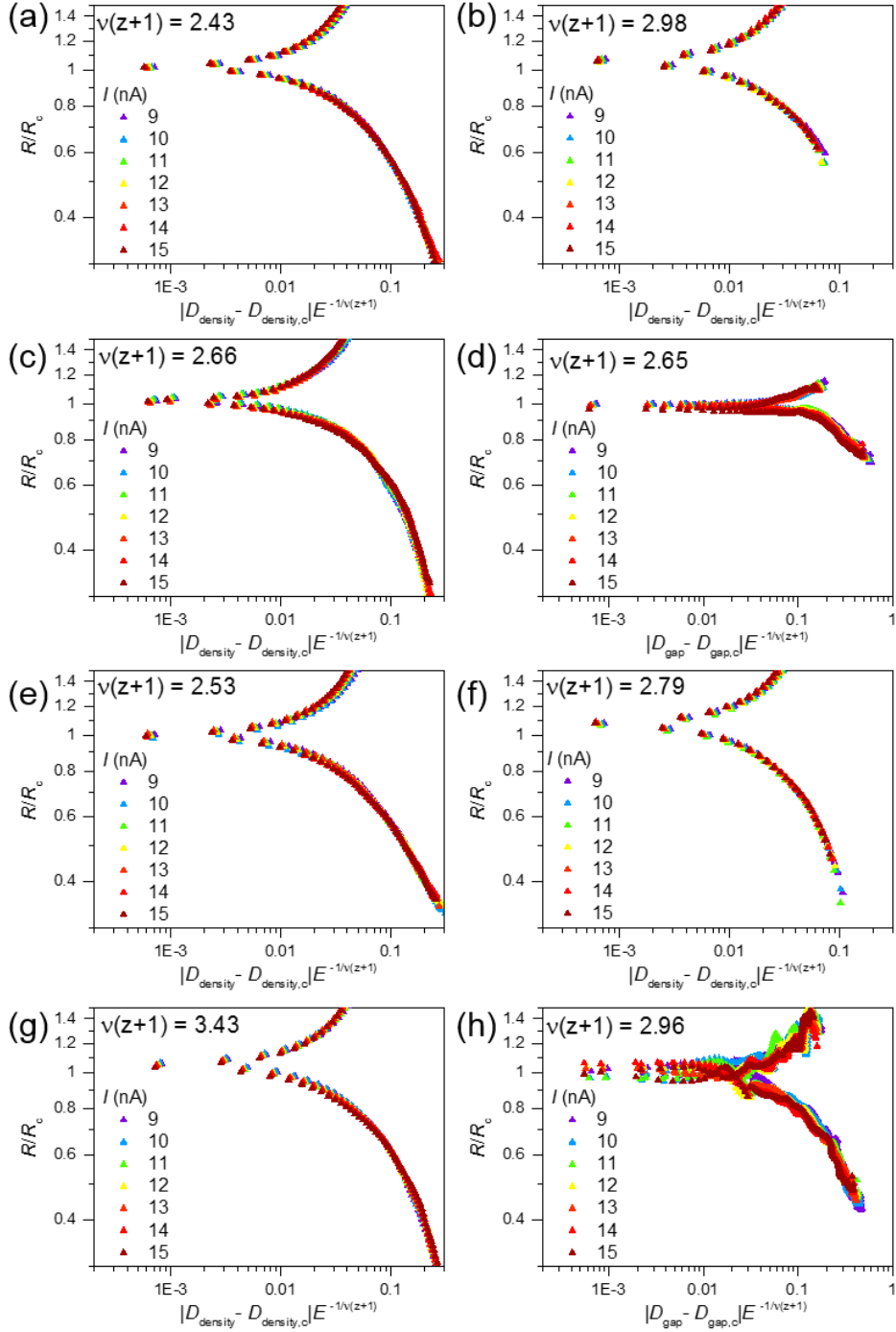

**Fig. S8.** Scaling analysis for electric-field variation. For the sweep 1 (a) in the hole and (b) electron sides. For (c) the sweep 2 and (d) 3. For the sweep S1 (e) in the hole and (f) electron sides. For (g) the sweep S2 and (h) S3.

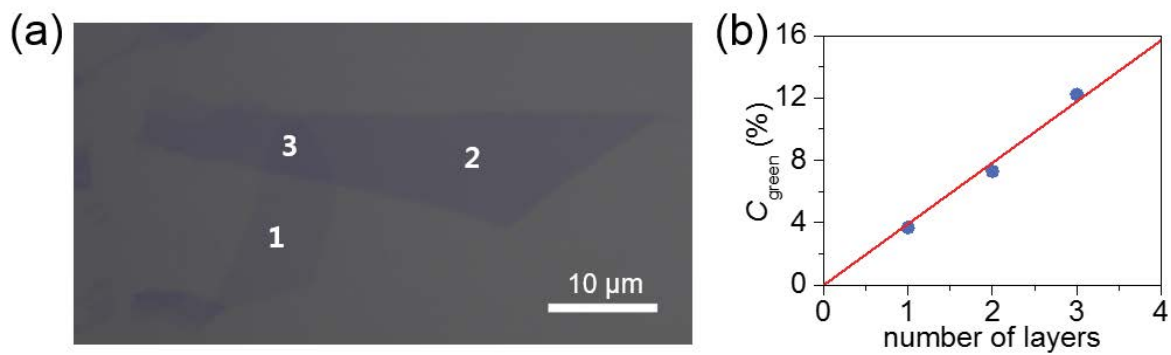

**Fig. S9.** Determination of the number of graphene layers. (a) Optical image of the graphene flakes exfoliated on an oxidised silicon substrate. The number of the graphene layers is denoted. (b) Linear relationship between the green-light contrast and the number of graphene layers. Red line crossing the origin is the best linear fit to the data.